\documentclass[12pt]{article}
\usepackage{authblk}
\usepackage[bookmarksnumbered, colorlinks, plainpages]{hyperref}
\usepackage{amsmath, amsthm, amscd, amsfonts, amssymb, graphicx, color, booktabs}
\numberwithin{equation}{section}
\definecolor{email}{rgb}{0.00,0.00,0.84}
\begin{document}
\setcounter{page}{1}

\title{\large \bf 12th Workshop on the CKM Unitarity Triangle\\ Santiago de Compostela, 18-22 September 2023 \\ \vspace{0.3cm}
\LARGE Inputs for the $\gamma$ measurements from BESIII}

\author[1]{Xiaokang Zhou on behalf of the BESIII Collaboration}
\affil[1]{Central China Normal University, Institute of Particle Physics, Wuhan, China}
\maketitle

\begin{abstract}
The CKM angle $\gamma$ is important for testing the unitarity of the CKM matrix and searching for new physics. $\gamma$ can be extracted by the interference between $b\to u$ and $b\to c$ in the B factory such as LHCb and Belle-II. Determining $\gamma$ also needs strong parameter information from the charm factory, such as the BESIII experiment. With quantum-correlated data samples collected at BESIII, the $\gamma$ uncertainties from the charm sector can be highly suppressed. 
\end{abstract} \maketitle

\section{Introduction}

\noindent Study of CP violation is one of the central topics in flavour physics. CP violation effects in the Standard Model (SM) can be described by one single complex phase angle in the Cabibbo-Kobayashi-Maskawa (CKM) matrix\cite{CKM1,CKM2}. Unitarity of the CKM matrix formulates the equation usually: 
\begin{equation}\label{eq_unitarity trignle}
\begin{aligned}
V_{ud}V^{\ast}_{ub}+V_{cd}V^{\ast}_{cb}+V_{td}V^{\ast}_{tb}=0.
\end{aligned}
\end{equation}
This equation can be described by a unitary triangle on the complex plane, where $\gamma$ (also known as $\phi_3$) is defined as:
\begin{equation}\label{eq_CKMangle}
\begin{aligned}
\gamma&=\phi_3=\left(-\frac{V_{ud}V^{\ast}_{ub}}{V_{cd}V^{\ast}_{cb}}\right).
\end{aligned}
\end{equation}
Precision measurement of the angle $\gamma$ is essential to testing the unitarity of the CKM matrix, and searching for the traces of new physics beyond the SM.

\subsection{$\gamma$ in $B$ factory}
$\gamma$ can be measured directly via tree-level decay, which is theoretically clean ($\delta\gamma/\gamma<10^{-7}$)\cite{Brod:2013sga}. The ideal decay for measuring $\gamma$ is $B^-\to DK^-$ due to its large statistics and clean backgrounds. Considering the same final states modes $B^-\to D^0(\to f)K^-$ and $B^-\to\bar{D}^0(\to f)K^-$, we obtained:
\begin{equation}
\begin{aligned}
A(B^-\to(f)_{\bar{D^0}}K^-)/A(B^-\to(f)_{D^0}K^-)=r_Br_De^{i(\delta_B-\gamma+\delta_D)},
\end{aligned}
\end{equation}
where $r_B(r_D)$ and $\delta_B(\delta_D)$ are the amplitude ratio and strong phase of $B(D)$ decays. The decay rate can be written as:
\begin{equation}
\begin{aligned}
\Gamma(B^{\mp}\to(f)_{D^0}K^{\mp})\propto1+r_B^2r_D^2+2r_Br_D\mathcal{R}\cos(\delta_B\mp\gamma-\delta_D),\\
\Gamma(B^{\mp}\to(f)_{\bar{D}^0}K^{\mp})\propto r_B^2+r_D^2+2r_Br_D\mathcal{R}\cos(\delta_B\mp\gamma+\delta_D),
\end{aligned}
\end{equation}
where $\mathcal{R}$ represents an average factor. Since the decay rates are correlated to observed signal yields, $\gamma$ can be extracted by combining the equations above. It can be easily found that to measure $\gamma$, we also need the strong parameters information from charm decays. 

Several methods can be used to extract $\gamma$ according to different $D$ decays:

If the neutral $D$ mesons decay to (quasi-)CP final states such as $K^+K^-$, $\pi^+\pi^-$, $\pi^+\pi^-\pi^0$, et al. This method is called ``GLW'' method\cite{Gronau:1991dp,Gronau:1990ra}. In this case, $r_D=1$ and $\delta_D=0$ or $\pi$ and $\mathcal{R}=2F_+-1$, where $F_+$ is called ``CP even content''(for pure CP even mode, $F_+=1$), which is the CP even purity for the modes and can be provided by charm factory such as BESIII, CLEO-c.

If the neutral $D$ mesons decay to Cabibbo-Favored(CF) and Doubly-Cabibbo-Suppressed(DCS) modes, such as $K^{\pm}\pi^{\mp}$, $K^{\pm}\pi^{\mp}\pi^{0}$, $K^{\pm}3\pi^{\mp}$ and so on. This method is called ``ADS'' method~\cite{Atwood:1996ci,Atwood:2000ck}. In this case, $r_D/\delta_D$ is unknown and needs to be provided from the charm factory. For multiply final states decay such as  $K^{\pm}\pi^{\mp}\pi^{0}$, $K^{\pm}3\pi^{\mp}$, $\mathcal{R}$ is the coherence factor, which represents the average effect of the phase space.

The most commonly used method is the so-called ``BPGGSZ'' method~\cite{Giri:2003ty,Bondar:2008hh}, which uses the $K_S\pi^+\pi^-/K_SK^+K^-$ modes, with large branching fractions and high reconstruction efficiencies. Considering the rich resonances in the Dalitz plane, which will provide more information for $\gamma$. One can divide the Dalitz plane into several bins accordingly, then combine the information in each bin to extract $\gamma$. In the ``BPGGSZ'' method, one also needs to know the charm strong parameters $c_i/s_i$ in bin $i$,  which $c_i=r_D\cos\delta_D^i$ and $s_i=r_D\sin\delta_D^i$. 

The latest world average result for $\gamma$ is $(65.9^{+3.3}_{-3.5})^{\circ}$~\cite{HFLAV}, which is dominated by the LHCb measurement ($\gamma=(63.8^{+3.5}_{-3.7})^{\circ}$)~\cite{LHCb:2022awq}. The most precise single measurement for $\gamma$ to date is $\gamma=(68.7^{+5.2}_{-5.1})^{\circ}$~\cite{LHCb:2020yot}, which is from ``BPGGSZ'' method. The uncertainty from the $c_i/s_i$ input is about $1^{\circ}$, which is dominated by the BESIII measurement based on the dataset of about 2.9~fb$^{-1}$ threshold charm data collected in 2010 and 2011~\cite{BESIII:2020hlg,BESIII:2020khq}. 

\subsection{Strong parameters in charm sector}
In the charm factory, the wave function for the $e^+e^-\to\psi(3770)\to D^0\bar{D^0}$ is anti-symmetric. 
This provides us with a unique opportunity to measure the strong parameters by using a double-tag(DT) method. 
We define the following quantities for the final state $i$\cite{CLEO:2012fel}:
\begin{equation}
\begin{aligned}
r^2_i=\frac{\int\bar{A}_i(x)\bar{A}^*_i(x)dx}{\int A_i(x)A^*_i(x)dx},\\
R_ie^{-i\delta_i}=\frac{\int\bar{A}_i(x)\bar{A}^*_i(x)dx}{r_iA^2_i},
\end{aligned}
\end{equation}
where $A_i=\langle i|D^0\rangle$, $\bar{A}_i=\langle i|\bar{D}^0\rangle$, and the integral is over the phase space for mode $i$. Thus, $\delta_i$ is an average phase for the final state $i$, and $\mathcal{R}_i\in[0,1]$. If the final state is two-body, such as $K^-\pi^+$, then it occupies a single phase space point, and $\mathcal{R} = 1$.

For a $D^0\bar{D}^0$ pair produced through the $\psi(3770)$ resonance, the decay rate to an exclusive final state {$i,j$}, where $i$ and $j$ are the final states of the two $D$ mesons can be denoted as:
\begin{equation}
\begin{aligned}
\Gamma(i,\bar{j})=\Gamma(\bar{i},j)\propto A^2_iA^2_j(1+r^2_ir^2_j-2r_i\mathcal{R}_i\cos\delta_ir_j\mathcal{R}_j\cos\delta_j-2r_i\mathcal{R}_i\sin\delta_ir_j\mathcal{R}_j\sin\delta_j),\\
\Gamma(i,j)=\Gamma(\bar{i},\bar{j})\propto A^2_iA^2_j(r^2_i+r^2_j-2r_i\mathcal{R}_i\cos\delta_ir_j\mathcal{R}_j\cos\delta_j+2r_i\mathcal{R}_i\sin\delta_ir_j\mathcal{R}_j\sin\delta_j),
\end{aligned}
\end{equation}
where the latter rate is reduced by half if $i$ and $j$ are identical. The above amplitudes are normalized to the uncorrelated branching fractions $\mathcal{B}_i$:
\begin{equation}
\begin{aligned}
\mathcal{B}_i=\mathcal{B}(D^0\to i)=A^2_i[1+r_i\mathcal{R}_i(y\cos\delta_i+x\sin\delta_i)],\\
\mathcal{B}_{\bar{i}}=\mathcal{B}(\bar{D}^0\to i)=A^2_i[r^2_i+r_i\mathcal{R}_i(y\cos\delta_i-x\sin\delta_i)].
\end{aligned}
\end{equation}
which correspond to the CF and DCS modes. 

Notice when $i$ is a (quasi-)CP eigenstate, then $\mathcal{R}=2F_+-1$, the uncorrelated branching fraction become:
\begin{equation}
\begin{aligned}
\mathcal{B}_i=A^2_i[1-(2F_+-1)y].
\end{aligned}
\end{equation}

From these equations above, we can extract the strong parameters by combining the information of different tags.

\section{Recent measurements of strong parameters}

\subsection{Quantum correlated charm dataset}

The BESIII experiment has collected a quantum-correlated $D^0\bar{D}^0$ data sample from $\psi$(3770) decay with an integrated luminosity of about 2.9~fb$^{-1}$ in 2011 and 2012. From 2022 to 2024, another 17~fb$^{-1}$ data has been collected and will be ready for physics analysis soon. In this paper, all the results are based on the 2.9~fb$^{-1}$ data.

\subsection{Improved measurement for $\delta_{K\pi}$}
\begin{figure} [hbt!]
\centering
\includegraphics[width=0.45\textwidth]{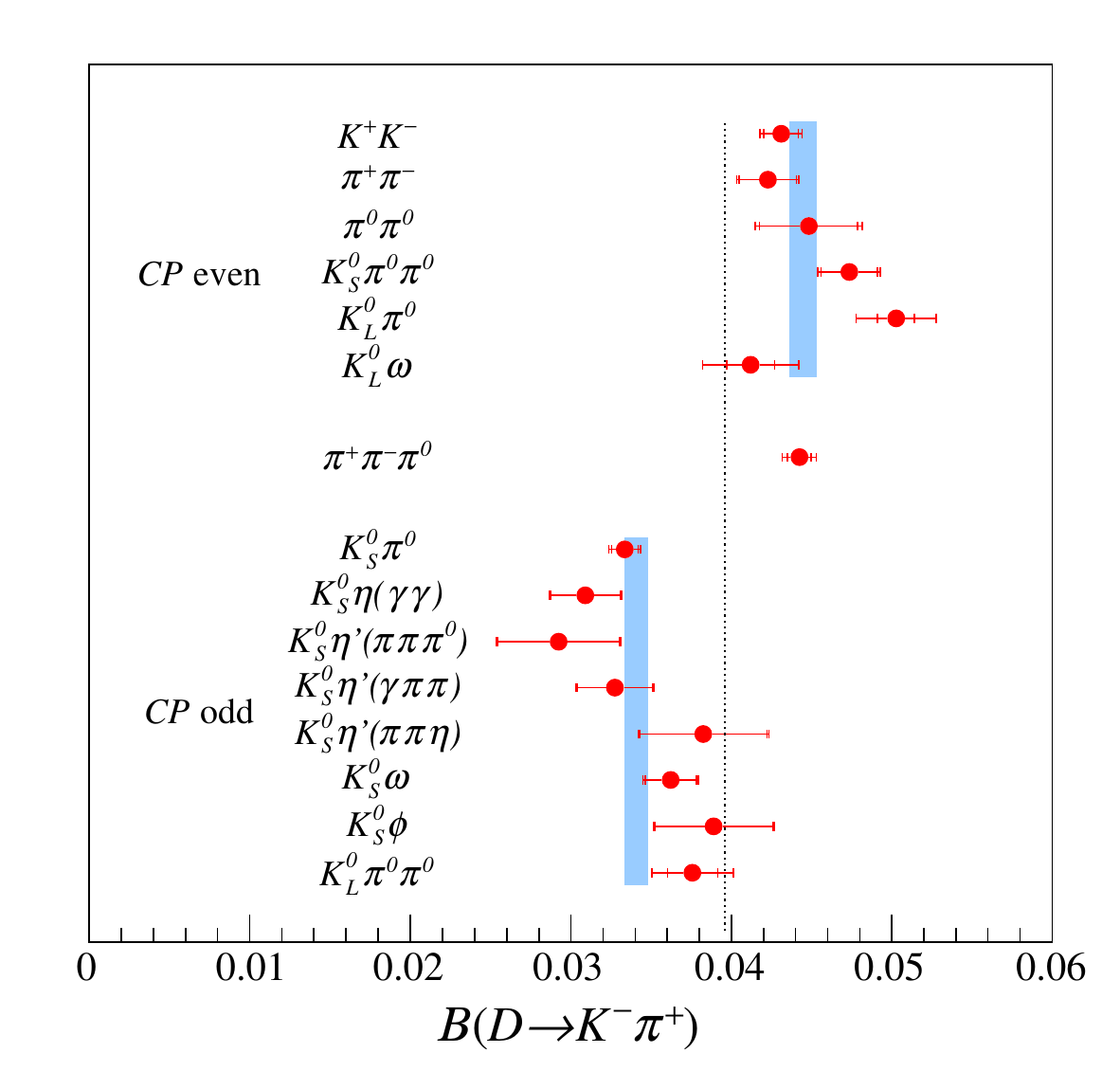}
 \caption{Branching fractions of $D_-\to K\pi$, $D_x\to K\pi$ and $D_+\to K\pi$ determined with different tag modes. The blue bands indicate the averaged result, and the vertical dotted line shows the measured central value.}
\label{fig:B_kpi_draft}
\end{figure}

The strong phase of $D\to K\pi$ mode is the important input for $\gamma$ measurement in the ADS method. In 2014, the BESIII collaboration reported the asymmetry between CP-odd and CP-even eigenstate decays into $K\pi$ to be $A_{K\pi}=0.127\pm0.013\pm0.007$~\cite{BESIII:2014rtm}, which only based on CP tag modes. An improved measurement is done with more tag modes included, such as $D\to K_LX$, here $X$ denotes the $\pi^0/\eta/\omega$, et al. These decay modes with $K_L$ improve the statistics of the CP tags a lot. Fig~\ref{fig:B_kpi_draft} shows the branching fraction results for $D\to K\pi$ in different CP tag modes. 

According to the observed yields for different CP tags, we can obtain the asymmetry $A_{K\pi}=0.132\pm0.011\pm0.007$~\cite{BESIII:2022qkh}, and $A_{K\pi}$ is related to $\delta_{K\pi}$ with 
\begin{equation}
\begin{aligned}
A_{K\pi}=\frac{-2r_{K\pi}\cos\delta_{K\pi}+y}{1+r^2_{K\pi}}.
\end{aligned}
\end{equation}

\begin{figure} [hbt!]
\centering
\includegraphics[width=0.48\textwidth]{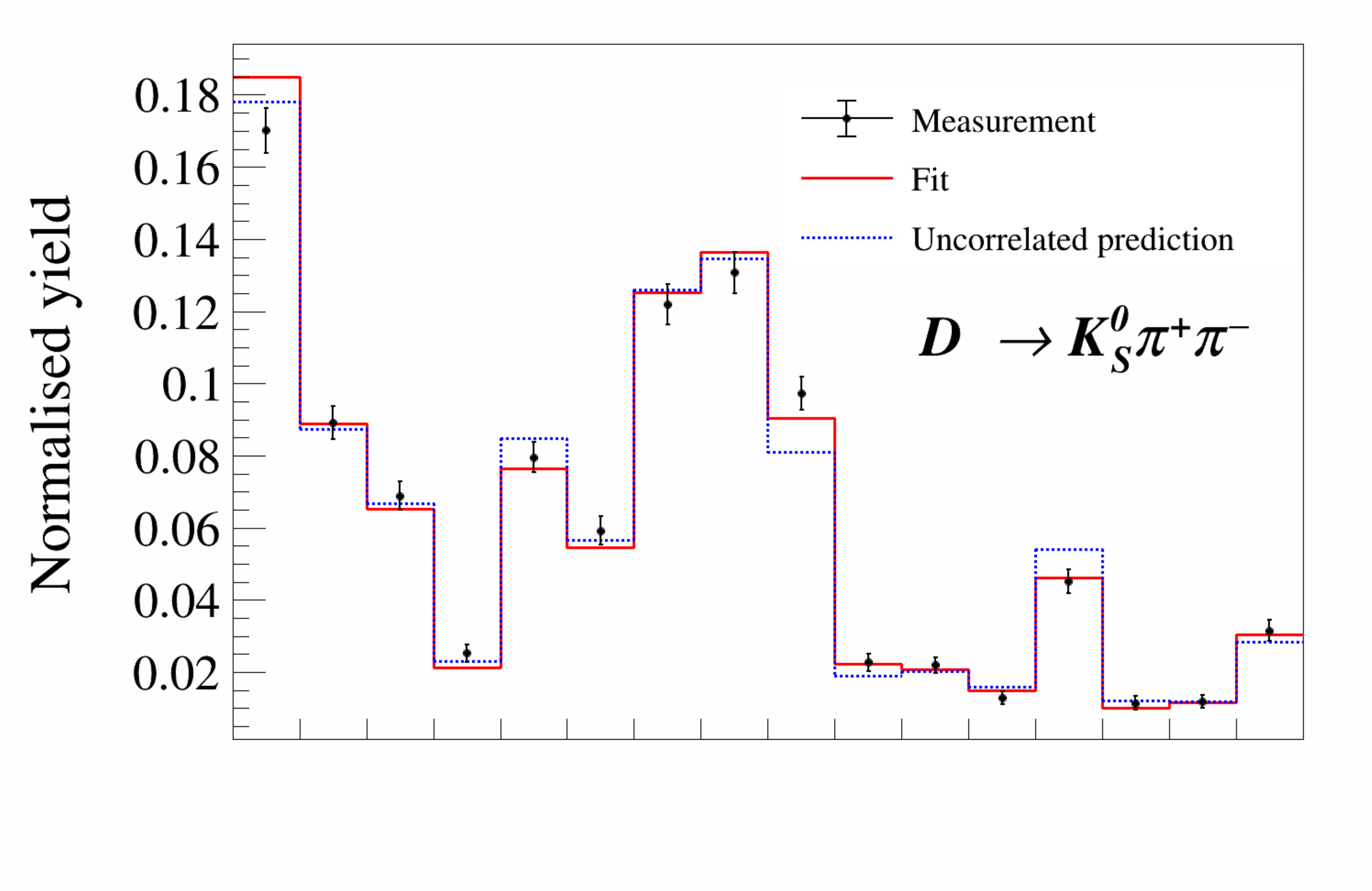}
\includegraphics[width=0.48\textwidth]{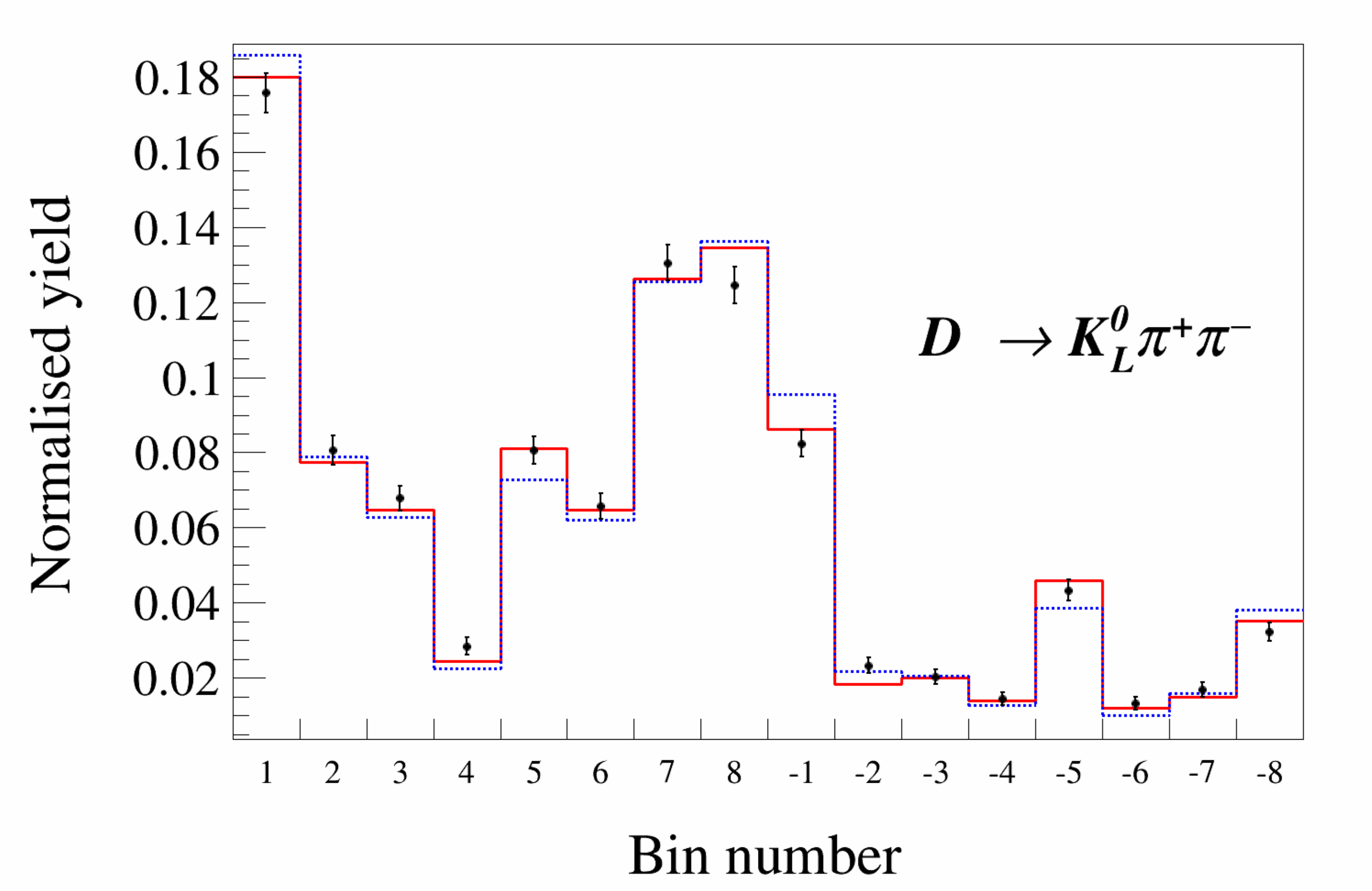}
 \caption{Fits to the $D\to K\pi$ sample tagged with $D\to K_S\pi\pi$ decays and $D\to K_L\pi\pi$ decays.}
\label{fig:binfit1}
\end{figure}

To further suppress the statistical uncertainties, the $D\to K_S/K_L\pi\pi$ modes are also included. We follow the ``equal $\Delta\delta_D$ binning scheme'' for $D\to K_S/K_L\pi\pi$ in the Dalitz plane and the yields in each bin have:
\begin{equation}
\begin{aligned}
N_i\propto[K_i+r^2_{K\pi}K_{-i}-2r_{K\pi}\sqrt{K_iK_{-i}}(c_i\cos\delta_{K\pi}-s_i\sin\delta_{K\pi})],
\end{aligned}
\end{equation}
where $K_i$ is the flavor-tagged fraction for $D\to K_S/K_L\pi\pi$, $c_i$ and $s_i$ are the strong parameters of $D\to K_S/K_L\pi\pi$. All these parameters have been measured by BESIII for $D\to K_S/K_L\pi\pi$ decays~\cite{BESIII:2020hlg,BESIII:2020khq}, and are used as input parameters here. A $\chi^2$ fit is performed to the normalised yields in the 32 phase-space bins of the two tagging modes, as Fig~\ref{fig:binfit1} shown.

Combing all the measurements above, and with some external parameters input, the strong phase difference for $D\to K\pi$ mode is determined to be $\delta_{K\pi}=(187.6^{+8.9+5.4}_{-9.7-6.4})^{\circ}$, where the first is the statistic uncertainty and the second systematic, respectively. This is the most precise measurement of $\delta_{K\pi}$ to date. 
The branching fractions are also determined:
\begin{equation}
\begin{aligned}
\mathcal{B}_{D^0\to K_L\pi^0}=(0.97\pm0.03\pm0.02)\%,\\
\mathcal{B}_{D^0\to K_L\omega}=(1.09\pm0.06\pm0.03)\%,\\
\mathcal{B}_{D^0\to K_L\pi^0\pi^0}=(1.26\pm0.05\pm0.03)\%. \nonumber
\end{aligned}
\end{equation}

\subsection{Measurement of the CP-even fraction}

As mentioned before, the CP-even fraction is an important external input for $\gamma$ measurement using the GLW method. By using different tag modes, one can extract the corresponding CP-even fraction in BESIII by $F_+=N_+/(N_++N_-)$, where $N_+$ and $N_-$ are the normalized yields of CP-odd and CP-even tags. Besides these CP tag modes, the $D\to K_S/K_L\pi\pi$ modes are also considered to improve the statistics, yields in each bin have
\begin{equation}
\begin{aligned}
N_i\propto[K_i+K_{-i}-2\sqrt{K_iK_{-i}}c_i(2F_+-1)].
\end{aligned}
\end{equation}
Then the $F_+$ can be determined.

With 2.9 fb$^{-1}$ of $e^+e^-\to\psi(3770)\to D^0\bar{D^0}$ data collected by
the BESIII experiment, the CP-even fractions are determined~\cite{BESIII:2022wqs,BESIII:2022ebh,BESIII:2023xgh}: 
\begin{equation}
\begin{aligned}
F_+^{\pi^+\pi^-\pi^+\pi^-}=0.735\pm0.015\pm0.005,\\
F_+^{K^+K^-\pi^+\pi^-}=0.730\pm0.037\pm0.021,\\
F_+^{K_S\pi^+\pi^-\pi^0}=0.235\pm0.010\pm0.002, \nonumber
\end{aligned}
\end{equation}
where the first is the statistic uncertainty and the second systematic, respectively. All these are the most precise measurements of the corresponding CP-even fraction to date. Some have been used to measure the $\gamma$. With more data collected by BESIII, these measurements will be significantly improved.

\section{Summary}
The measurements of $\gamma$ are still statistically limited to date. However, LHCb began the new round of data taking, Belle-II is also on data taking. The systematic uncertainty of $\gamma$ will become the dominant uncertainty, especially for these systematics from charm inputs. The 20~fb$^{-1}$ of $D^0\bar{D^0}$ data collected by BESIII experiment will be soon ready, which  will reduce these uncertainties a lot and help us to test the SM in high precision.

\bibliographystyle{amsplain}

\end{document}